\documentclass[acmnone,acmnow]{acmtransmod}

\usepackage[usenames,dvipsnames]{color}

\usepackage{times}
\renewcommand{\baselinestretch}{1.1}
\usepackage{url}
\usepackage{verbatim}
\usepackage{pdfpages}

\usepackage{graphicx}
\DeclareGraphicsExtensions{.ps,.pdf,.png,.tif,.jpg}
\usepackage{subfigure}
\usepackage{wrapfig}

\usepackage{mathptmx}
\usepackage{amsmath,amsfonts,amssymb,euscript,mathrsfs,bbm}
\usepackage{theorem}

\newcommand{\sname}[1]{{\small \sf #1}} 

\newcommand \parao[1] {\vspace{.25em} {\setlength{\parindent}{0pt} \bf #1}}
\newcommand \para[1] {\parao{#1:}}

\newcommand{\SC}[1]{Section~\ref{#1}}

\newcommand{\EQ}[1]{Equation~\ref{#1}}

\newcommand{\FG}[1]{Figure~\ref{#1}}
\newcommand{\FIG}[1]{\FG{#1}}

\newcommand{\shah}{{\textstyle \amalg{\kern-4.pt\amalg}}}

\title{Mixing Board Versus Mouse Interaction In Value Adjustment Tasks}
\author{
	Steven Bergner, Matthew Crider, Arthur E. Kirkpatrick, Torsten M{\"o}ller \\\ %
    {\small School of Computing Science}, 
        {\small Simon Fraser University}, 
        {\small Burnaby, BC, Canada}
}

\usepackage[pdftex,bookmarks]{hyperref}
\hypersetup{
  pdfauthor = {Steven Bergner},
  pdftitle = {Mixing board vs. Mouse interaction in parameter adjustment tasks},
  pdfsubject = {},
  pdfkeywords = {Human-computer interaction, visualization},
  colorlinks=true,
  linkcolor= Blue,
  citecolor= black,
  pageanchor=true,
  urlcolor = black,
  plainpages = false,
  linktocpage
}

\begin{abstract} 

We present a controlled, quantitative study with $12$ participants comparing interaction with a haptically enhanced mixing board against interaction with a mouse 
in an abstract task that is motivated by several practical parameter space exploration settings.

The study participants received $24$ sets of
one to eight integer values between $0$ and $127$, which they had to match by making adjustments with
physical or graphical sliders, starting from a default position of $64$.
Based on recorded slider motion path data, we developed an analysis algorithm that identifies and measures different types of activity intervals, including {\em error} time moving irrelevant sliders and {\em end} time in breaks after completing each trial item. This decomposition facilitates data cleaning and more selective outlier removal, which is adequate for the small sample size.

Our results showed a significant increase in speed of the
mixing board interaction accompanied by reduced perceived cognitive load when compared with the traditional mouse-based GUI
interaction. We noticed that the gains in speed are largely due to the improved times required for the hand to reach for the first slider ({\em acquisition} time) and also when moving {\em between} different ones, while the actual time spent {\em manipulating} task-relevant sliders is very similar for either input device. 
These results agree strongly with qualitative predictions from Fitts' Law that the larger targets afforded by the mixer handles contributed to its faster performance.

Our study confirmed that the advantage of the mixing board acquisition times and between times increases, 
as more sliders need to be manipulated in parallel.
To investigate this further, we computed a measure of motion simultaneity based on velocity correlation. This enabled us to identify types of items for which increased simultaneous adjustments occur.

In the context of continuous parameter space exploration our findings suggest that mixing boards are a considerable option that provides a detailed multi-value control. The strengths of this input method should particularly show in settings where screen space is precious and undisrupted visual focus is crucial.

\end{abstract}

\category{H.5.2}{Information interfaces and presentation (e.g., HCI)}{\\User Interfaces, Input devices and strategies (e.g., mouse, touchscreen)}

\keywords{mixing board, tangible interaction techniques, TUI}

\begin{document}


\maketitle

\section{Introduction}
Many applications require frequent adjustment of numeric or categorical parameters.  In audio and video production, properties of the sound or video tracks must be adjusted frequently.   In visualization, applications transform data into interactive images that can be explored by adjusting parameter values.  

Traditionally, audio mixing has been performed using large arrays of analog sliders.  These devices are so closely associated with that task that they are commonly referred to as ``mixing boards'' or ``mixers''.  With the shift to digital media, the underlying mechanics have become digital, but the physical interface remains an array of sliders.

From the perspective of interface style, a mixing board occupies an
interesting point at the intersection of tangible user interfaces,
augmented reality, and haptic user interfaces.  The sliders embody the
benefits of a tangible interface: their handles engage the whole hand
and provide strong feedback on contact, they are physically
constrained to a single direction of motion, they have physical
backstops stopping motion at their limits of travel, and each slider
is maintained in fixed physical relationship relative to the
others. Augmenting the mixing board by front-projecting graphical
output on it~\cite{Crider:2007:MBI}, integrates the space of the
user's movements with the display space, a form of augmented reality.
Finally, some mixing boards have motorized sliders, allowing software
controlled haptic effects such as detents.

Although a mixing board offers the benefits of these three interface styles, it is also limited compared to more typical members.  Movement and display of a slider are restricted to a single degree of freedom, with all sliders in the same plane and direction, whereas most tangible and augmented reality interfaces emphasize movement in three or six degrees of freedom.  Yet restricted as it is, the interaction task supported by the mixing board, entering several bounded numeric values, recurs in applications ranging from scientific and medical data analysis, computer-aided design, to setting document margins. 

In a prior qualitative study, participants reported a greater sense of
engagement and productivity when using the mixing board for such
tasks~\cite{Crider:2007:MBI}.  In this study, we focus on quantifying
the performance benefits from the tangible properties of the mixing
board.  We compare its acquisition time, movement time, and workload
demands to those of a similar array of graphical sliders.  We
also compare subjective, qualitative evaluations of the two interfaces
by our 12 participants.

\section{Previous Work}

Specialized, spatially-multiplexed input devices have been shown to outperform generic, temporally-multiplexed devices (such as mice) for a variety of applications.  Fitzmaurice and Buxton~\cite{Fitzmaurice97} demonstrated that physical or ``graspable'' user interfaces with specialized shapes and dedicated functions were superior to a generic input device for a target tracking task. In particular, they found only half the time with the mouse-based interface was spent actually adjusting the graphical widgets with the mouse, while for the specialized graspable interface,  about 89\% of the time was spent moving the physical devices.  The difference between the two was largely due to the greater time required to select the graphical widget with the mouse over grabbing a physical device with the hand.

Temporally-multiplexed input devices such as mice have a potential cognitive load as well.   They require users to first acquire the device, move the pointer to acquire a graphical widget, and then manipulate the widget. As a result, such devices can break the flow of a user's cognitive engagement with the task, negatively impacting performance \cite{Faisal05}. Jacob and Sibert's work~\cite{Jacob:1994:integral} demonstrated that input devices which support the type of motion required by the task may alleviate this problem. Additionally,
people may be able to use physical devices without removing their
gaze from their object of interest, unlike graphical widgets which require that 
direct visual focus be shifted to them and away from the object of interest.
Such non-visual interaction may promote
better engagement with the task and more efficient performance. 

Several studies have specifically compared various forms of sliders. Hunt and Kirk~\cite{HuntK99} 
compared physical and virtual sliders for setting parameters in a sound matching task.
By using a computer and human marker to gauge how close they were to the correct sounds, they determined that their participants achieved better results
using physical sliders. Chipman et al. \cite{chipman04}
compared a physical slider, a graphical scrollbar, and the mouse
wheel for two scrolling tasks. Both physical interfaces performed
better than the graphical scrollbar, with the mouse wheel being
superior for searching and the physical slider being superior for a
reciprocal tapping task.  

Mixing boards have long been used to control input and
output signals in video and audio applications. However, mixing boards have rarely been used in other classes of applications. Shahrokni et. al. \cite{Shahrokni2006}
applied a force-feedback slider in visualizing laws of
physics in an educational game, but their work is only a proof-of-concept and offers no analysis of the device. Rheingans~\cite{Rheingans92} reported using physical sliders to
interactively control color mappings in a visualization tool, although
the physical interface was not the focus of her work.

A few customized physical interfaces have been developed for spatial
navigation in 3D visualization applications (e.g., Hinckley et
al.~\cite{Hinckley94} and Konieczny et
al.~\cite{KoniecznySMC05}). Much less work has been done to examine
physical interfaces for non-spatial controls.
SeismoSpin~\cite{McKelvin03}, a custom physical input device for
interacting with earthquake visualizations, was reported to be more
engaging and interactive than traditional interfaces. 
 
Tangible controls also permit two-handed input, which can be more efficient for at least two reasons: less hand movement is needed because the two hands have resting positions on different controls, and for some types of input, hands can be used simultaneously~\cite{buxton86}.  
Kabbash et al.~\cite{Kabbash94} suggested that the two hands should be used in a dependent way,
with the non-dominant hand setting a frame of reference for precise action performed by the dominant hand. We do not suggest that our mixing board is an optimal two-handed interface, but the ability to use two hands simultaneously may impart some benefit.

\para{Research Questions} 
The goal of our study was to answer several questions regarding the effectiveness of a mixing board as input device, and not necessarily in the context of a specific application.  By measuring the time it takes to complete certain tasks with either the mixing board or a mouse, observing our study participants, and asking them about their experience, we hoped to determine:
\begin{itemize}
  \item Which device, overall, completes the given tasks more quickly and why?
  \item Which types of tasks does the mixing board excel in?
\item The mixer allows for simultaneous adjustments. For what types of items do they occur?
\item Do target values that are near or far from the initial position for different windows of precision cause a difference in performance (confirming Fitts' Law)?
  \item What adjustment strategies do participants follow to accomplish their tasks?
  \item Which device do participants subjectively prefer, and for what reasons?
\end{itemize}

\begin{figure}[t]
\centering
\includegraphics[width=12cm]{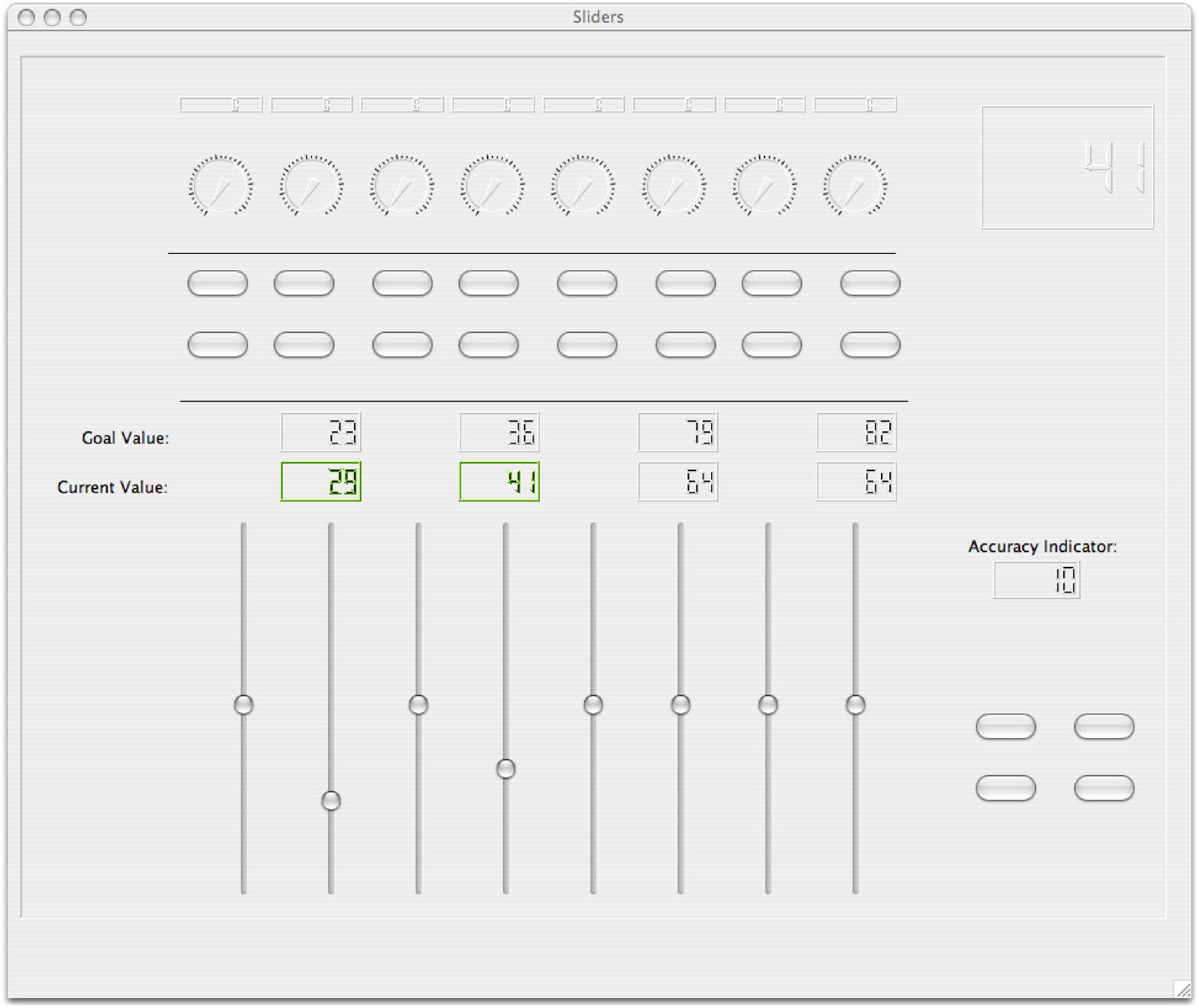}
\caption{Graphical user interface (GUI) for the BCF2000 showing an experimental trial in progress.
\label{fig:midTrial}}
\end{figure}
\section{Experimental Task and Study Design} \label{sec:task}

In order to cover different possible application scenarios, we chose to study an abstract multi-value adjustment task that is representative of different practical settings.
A description of the task and the study design is given in the following. Details of the hardware and software setup are postponed to Appendix~\ref{sec:setup}.

\para{Task}
In each trial, participants were presented a tuple of values and asked to set controlling sliders to those values within a specified precision, requiring one, two, four, or eight sliders to be set.  The screen displayed an interface (\FIG{fig:midTrial}) that resembles the BFC2000 mixing board used in the study (see \FIG{fig:bcf2000} below for comparison), except for one difference:    
The mixing board used a multiplexed display in its upper right corner to show current values of the last moving slider. The GUI displayed current and target values above each slider that the participant was asked to move in a trial. 
Adjustments to the graphical or the motorized physical sliders caused the corresponding movements on the alternate interface.

\para{Physical size}
The standard Mac OS X aqua sliders, in the initial study had a diameter of 5 mm on the screen. Moving the mouse pointer over this distance at a low velocity, amounted to about 8 mm of mouse movement.
The mixing board slider knobs measured $10\times16\times10$ mm. The length of a GUI slider was 75~mm and 110~mm for the mixer, spaced at 20 mm for the GUI and 27 mm on the board.

\subsection{Experimental Design \label{sec:expdes} }

The number of values to be set and the precision required were varied systematically 
using a six-factor, within-subjects design.  The primary factor was {\em technique} [mixer or GUI].  Order of technique was counterbalanced, with participants randomly assigned to one of the two possible orders.  The other five factors consisted of {\em number of sliders} [1, 2, 4, or 8], {\em precision} [loose $\pm7$ or tight $\pm1$], {\em distance} of target value [near ($\pm6$) the initial value, far ($\pm46$), or at a backstop (0 or 127)], and {\em slider layout} [adjacent or separated]. Post-hoc also {\em value layout} [aligned or opposing] could be distinguished among the items.
To keep the length of the experimental session manageable, these factors were only varied in a subset of the 192 possible combinations to create the sample targets---i.e. they were not fully-crossed.  
The resulting pool of 24 targets, listed in \FIG{fig:items}, was presented in a randomized order for every block for every participant.
\begin{figure}[ht]
\centering
\texttt{\scriptsize
\begin{tabular}{ll}
&{\small \bf 4 sliders:} \\
&\textnormal{\small 	Backstops or distance/precision combinations} \\
&\textnormal{\small 	various combinations of grouped versus} \\
   {\small \bf 1 slider:}  &\textnormal{\small 	combined and directions}\\
    \textnormal{\small Backstop or distance/precision combinations}  &10: [~~0~~~0~~~0~~~0~~~.~~~.~~~.~~~.~]+/- 1 \\
    1: [  .   .   .   .   .   .    0~~~. ]+/- 1&11: [~18~~18~~18~~18~~~.~~~.~~~.~~~.~]+/- 1\\
    2: [  .   .   .   .   .   .   58~~. ]+/- 1 &   12: [127~~~0~127~~~0~~~.~~~.~~~.~~~.~]+/- 1\\
    3: [  .   .   .   .   .   .  110  . ]+/- 1 &  13: [ 18  110~~18~110~~~.~~~.~~~.~~~.~]+/- 1\\
    4: [  .   .   .   .   .   .  110  . ]+/- 7 &    14: [ 18~~18~~18~~18~~~.~~~.~~~.~~~.~]+/- 7 \\
    {\small \bf 2 sliders:}  & 15: [ 18  110~~18~110~~~.~~~.~~~.~~~.~]+/- 7\\
    \textnormal{\small 	3 distance/precision combinations,} &16: [~~.~~~0~~~.~~~0~~~.~~~0~~~.~~~0~]+/- 1 \\
    \textnormal{\small 	and adjacent versus combined} &17: [~~.~127~~~.~~~0~~~.~127~~~.~~~0~]+/- 1\\
    5: [  .   .   .   .   .   .   ~70  ~70]+/- 1 &18: [~~.   18~~~.~~18~~~.~~18~~~.~~18~]+/- 1\\
    6: [  .   .   .   .   .   .  110 110]+/- 1  &19: [~~.~110~~~.~~18~~~.~110~~~.  18~]+/- 1\\
    7: [  .   .   .   .   .   .  110 110]+/- 7&20: [~~.   18~~~.~~18~~~.~~18~~~.  18~]+/- 7\\
    8: [  .   70  .   .   .   70  ~.   ~.~]+/- 1 &21: [~~.~110~~~.~~18~~~.~110~~~.  18~]+/- 7\\
    9: [  .~110  .   .   .~110  ~.   ~.~]+/- 7  & {\small \bf 8 sliders:} \\
 &	\textnormal{\small 	3 distance/precision combinations} \\
   & 22: [~70~~70~~70~~70~~70~~70~~70~~70~]+/- 1\\
    &23: [110~110~110~110~110~110~110~110~]+/- 1 \\
    &24: [110~110~110~110~110~110~110~110~]+/- 7\\
\end{tabular}
}
\caption{Set of 24 items used in our study design shown as tuples of slider target values.
In a trial, sliders needed to be adjusted to the given value configuration with a given tolerance of plus or minus one or seven. Values marked [~.~] indicate that that slider was not to be adjusted. Item~8 was run twice in a block, for a total of 24 trials/block.\label{fig:items}}
\end{figure}

\para{Collected data}
During a trial, the software recorded the changes to the digital slider values for either input device (128 discrete levels along 110 mm slider length on the mixer or 75 mm on the screen) as they occurred at a temporal resolution of at least 20 ms per measurement.
When sliders are moved using any input device, path data (as in in \FG{fig:pathdata}) is collected noting the change in value over time as one proceeds through the trials. 
\begin{figure}[ht]
\begin {center}
{\renewcommand{\baselinestretch}{.8}\small\begin{tabular}{ccc}
   {\bf timestamp (ms)}&   {\bf slider id} &    {\bf value } \\
         950&           2     &     65\\
        1052  &         2    &      64\\
        1094 &          1    &      63\\
        1114   &        1    &      62\\
        1483 &          4    &      63\\
        1503 &          3   &       63\\
        1503    &       4   &       62\\
        1524 &          3    &      62\\
        1524 &          4   &       61\\
        1546  &         3   &       61\\
        1546    &       4   &      60\\
        1567   &        1   &       61\\
        1567   &        2   &       63\\
\end{tabular}}
\caption{Slider path event recording from MIDI data. Example of the beginning of
simultaneous downward movement of 4 sliders as recorded for a trial of item 10 of \FG{fig:items}.\label{fig:pathdata}}
\end{center}
\end{figure}


The software recorded the final values of each slider and the trial start and end times, defined as the time the space bar was pressed. 
Dependent measures were {\em total} time (from start of trial to the end of movement of the last slider moved), and simultaneity of slider movement (for trials with multiple targets), 
and subjective workload, as measured by TLX~\cite{Hart:1988:TLX}.  

\subsection{Participants}

A convenience sample of 12 participants (8 male, 4 female) was recruited from graduate students (5 Ph.D., 7 MSc.) in computer science at Simon Fraser University. Their age ranged from 24 to 37 ({\em Mdn} 27).  All participants considered themselves to be right-handed. They were paid 20 CAD for their participation. 

\subsection{Experimental protocol}

Our participants did all trials in a single session.  They were first introduced to the methods that would be used in the study, then confirmed in writing their informed consent.  They next completed a questionnaire asking basic background information (age, gender, education, computer usage, and handedness).  
After a set of practice trials that was generated for each input device from the same pool of items as the main blocks, the participant performed three blocks of trials for their first technique, and then filled out a TLX questionnaire to estimate the workload for using the technique.
They next did three blocks of trials for their second technique, followed by a second TLX workload estimate.  Participants concluded the session by answering open-ended questions about the techniques and a question asking which technique they preferred. A typical session lasted about half an hour.

\para{Trial procedure}
Separately for each input method, the stimuli were presented in randomized order in three blocks of 24 target vectors of values.  During the mixer blocks, the board was placed in front of the screen.
Participants were allowed to rest as long as they wished between blocks. Trials within a block were immediately followed by the next, once their completion was confirmed by the participant pressing a mixer button or the space bar.
Participants were informed about the varying tolerated accuracy for the required adjustments. They were also told that a fixed set of trials lies ahead of them, indicating that a faster performance will reduce their time spent in the study.

A trial began when the participant pressed the space bar or the confirm button in the lower right of the mixing board.  The target values and the tolerated imprecision (accuracy indicator) would be displayed and all sliders were automatically initialized to their middle value (64) using the available motorized control.  
The participant would then move one or more sliders in whatever order they wished, setting them near their target values.  When a slider was within the required precision of its target value, the ``current value'' in the GUI display would turn green (e.g. two values in the left of \FIG{fig:midTrial}).  
The software only allowed to end a trial (using space bar or confirm button) when all targets were correctly specified. While this made error trials impossible, the time spent with superfluous adjustments could still be measured.

\section{Activity Interval Analysis and Outlier Detection} \label{sec:ivdecomp}

The slider motion path data is collected as a sequence of events of time stamped value updates, shown earlier in \FIG{fig:pathdata}. It can be graphed as in \FG{fig:ivs}, where also the result of the decomposition of the overall trial time is displayed. 
The six trials stem from two-slider item 6, recorded in three blocks using the mixing board (top row) and three blocks using mouse/GUI (bottom row). The gray intervals mark {\em idle} times when no valid slider was manipulated. The red sectors mark {\em error} times spent manipulating irrelevant sliders.

\begin{figure*}[hb]
\centering
\includegraphics[width=18cm]{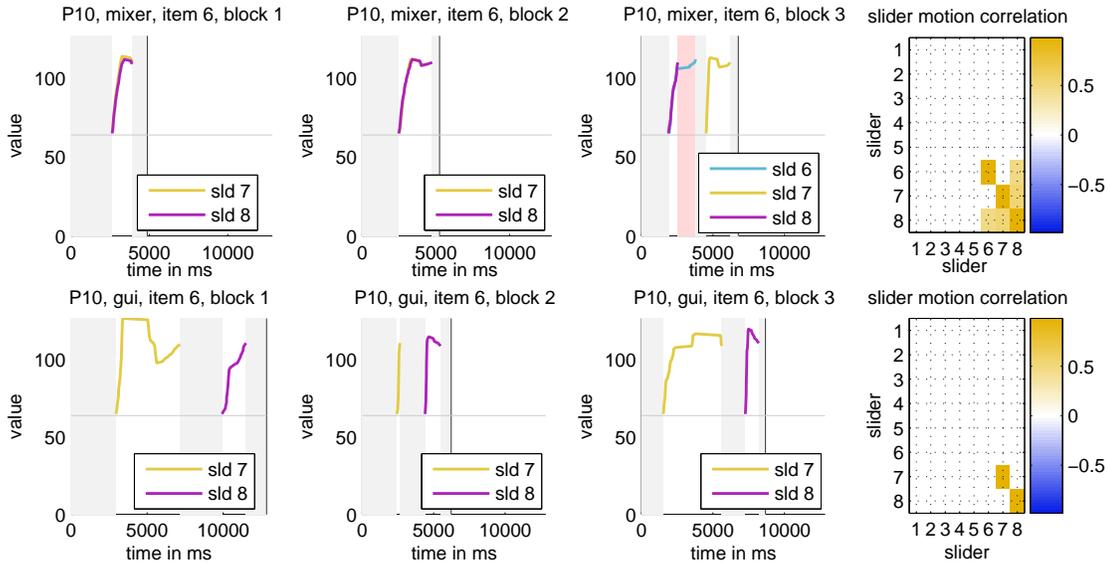}
\caption{\small Recorded slider motion paths for item {\small
\tt6:~[ .~.~.~.~.~.~110~110]+/-1} shown in front of non-manipulation
intervals (gray) and mistake intervals (red). The top row is mixer interaction of a participant in blocks
1-3 of item 6 of \FG{fig:items}, indicating an error moving irrelevant slider 6 in
the third block. The path correlation matrix in the right column is discussed in \SC{sec:pathcorr}.
\label{fig:ivs}}
\end{figure*}

\para{Types of activity intervals}
To clarify the detailed mechanics underlying the decomposition shown in \FG{fig:ivs}, movement time was broken down into several distinct subtimes, which were determined algorithmically using custom MATLAB scripts.  {\it Acquisition time} was computed as the time from the start of the presentation of a trial (end of the previous trial) to the movement of the first slider. For the mouse, it comprised the activities of possibly moving the hand from the space bar to the mouse (if the user ended the previous trial with their mouse hand), moving the mouse to the first slider, and clicking. For the mixer, acquisition time consisted of moving the hand from the trial end button on the right panel of the mixing board to the first slider.   {\it Manipulation time} was computed as the time any of the the sliders involved in the task were actually moving.  {\it Between time} was computed as the time spent between slider movements.  
{\it Total movement time} was computed as the time from the start of movement of the first slider to the end of movement of the last slider.  
For a given trial, the sum of acquisition, manipulation and between time equals the {\em total time}, not including the {\em end} time between the last slider movement and the pressing of the trial end button (space bar for the mouse, or a dedicated button on the mixer). 

We have also determined {\em error} time, which is the amount of time between {\em task relevant} slider movements when erroneous sliders have been moved. An example of such an event can be observed in the top row of \FG{fig:ivs} for the third block of the mixer interaction, in which slider 6 was moved in error. {\em Error time} is contained in the {\em between time} and is also reported separately.

\subsection{Outlier detection}
We have grouped the entire set of measurements
(150 trials for 12 participants) into 48 groups, 2 input devices per
each of the 24 items. Within each group we do a separate outlier
analysis. For that we have implemented two methods. One determines
mean and standard deviation per group (ignoring the first practice
block). Then all measurements that are outside a given radius of standard
deviations (z-score) are flagged as outliers. The second method chooses the
radius to be $\pm 4$ times center quartile range around the
median. A comparison of the two outlier detection methods is given in \FG{fig:outliers}. 
\begin{figure}[h]
\centering
\includegraphics[width=8cm]{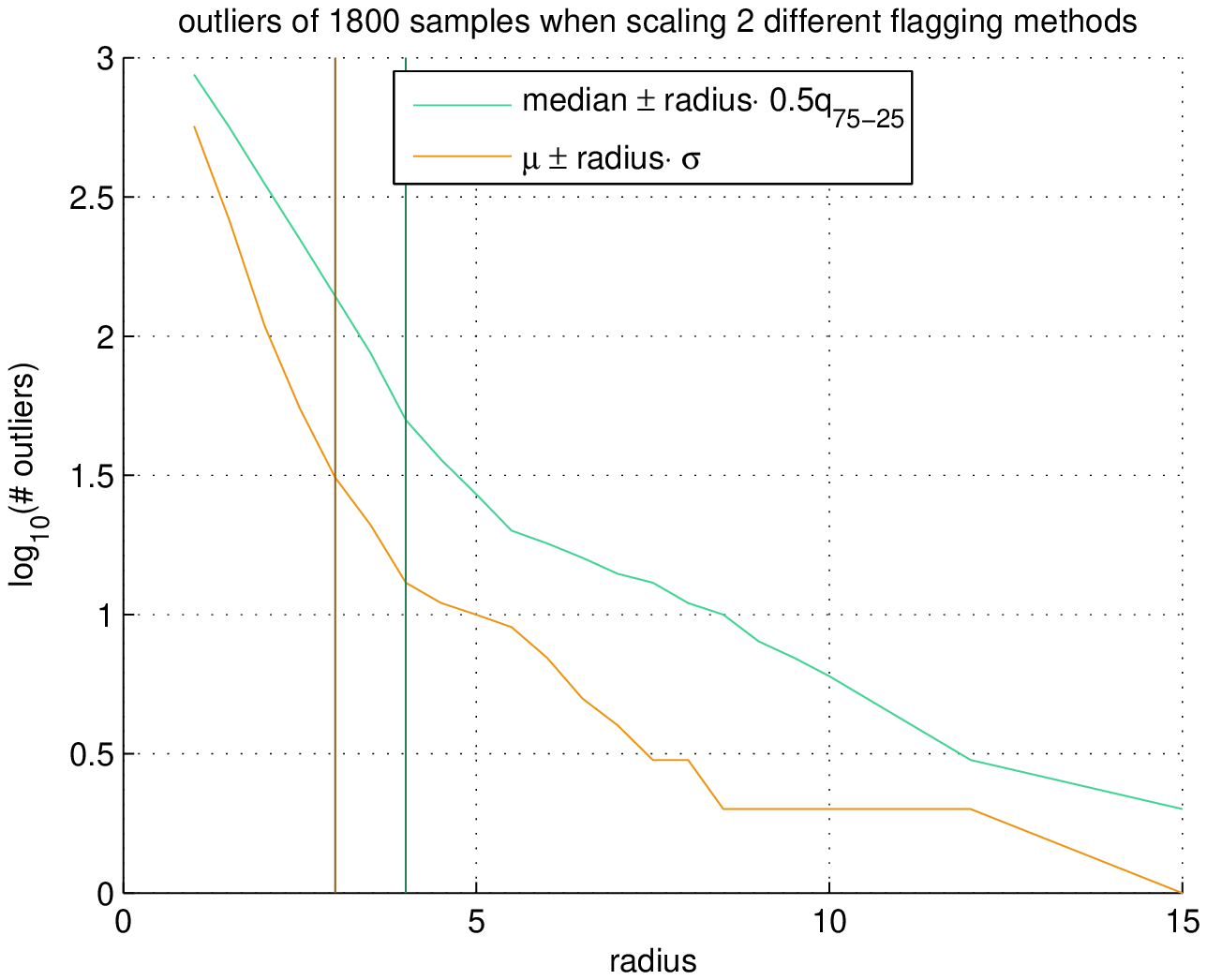}
\includegraphics[width=8cm]{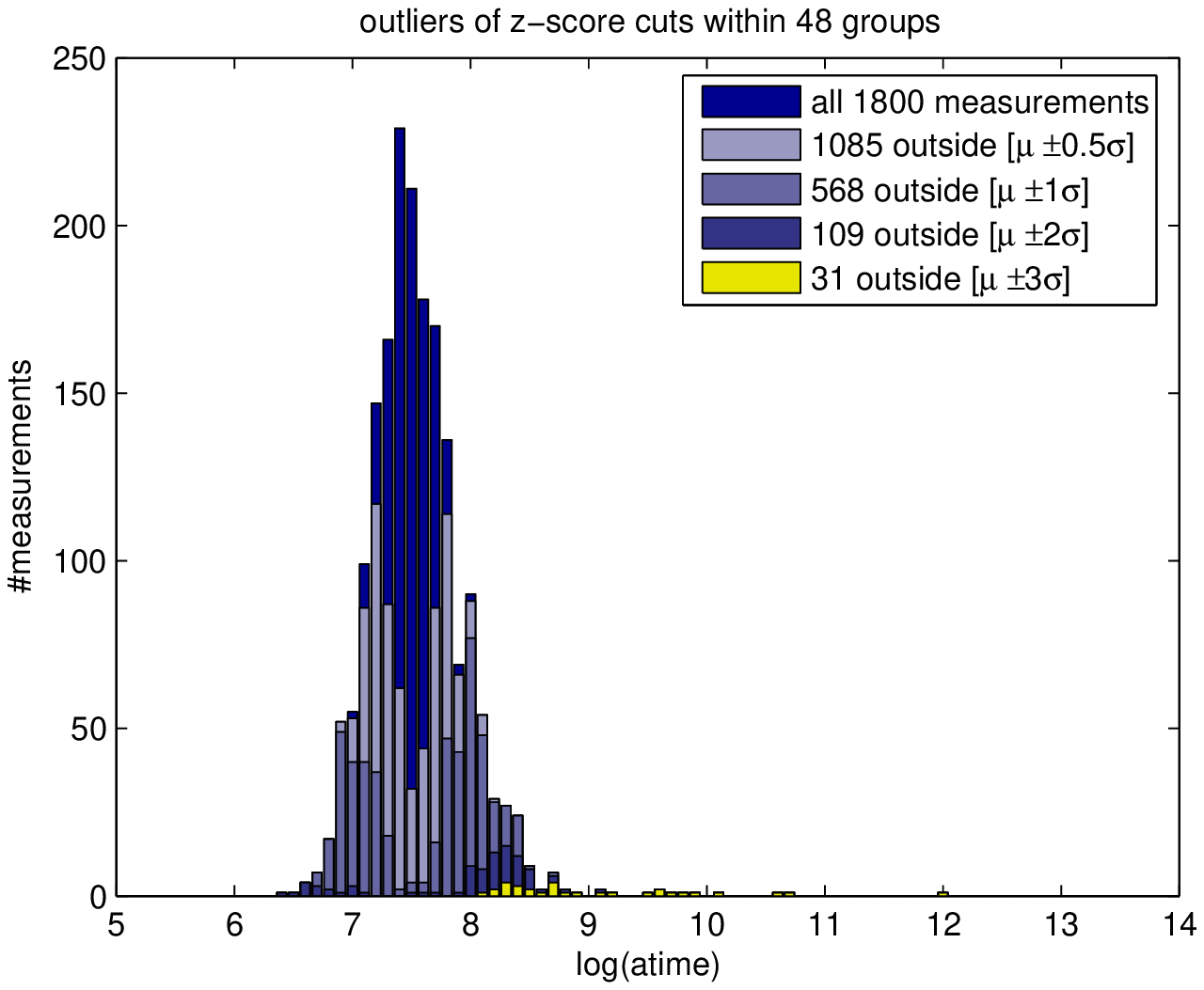}
\caption{(top) Outlier distribution when scaling two removal methods
  separately over 48 trial groups (2 devices $\times$ 24 items). (bottom) Outlier distribution for a z-score cut (orange line in top figure) at radius 3 (yellow) shown over trial acquisition time for the entire trial set. \label{fig:outliers}}
\end{figure}

Of the 31 flagged outliers 18 were due to pauses participants took at
the beginning of a trial and 9 were due to adjustment of wrong
sliders. The remaining ones resulted from interim testing of what effect moving a certain slider would have. Overall, 23 of the outliers were in practice block 1, which was excluded from final timing analyses. Outliers were computed in groups separately for each input device, flagging 24 for mouse/GUI input and 7 for the mixer.

\subsection{Motion path velocity correlation\label{sec:pathcorr}}
A key feature of the mixer interaction is the possibility of
simultaneous movement of sliders as the user may use the same hand or
both hands to affect multiple sliders at any given time of the
interaction. As a measure of simultaneous movement we have used path
velocity correlation. Slider velocities can be determined as the
time derivatives of positions $s_a(t)$ and $s_b(t)$,
e.g. $s'_a(t)=\frac{ds_a(t)}{dt}$. The time correlation of the movement
of the sliders is given via an integral over the products of
slider velocities over the duration $t=0..T$ of an entire trial
\begin{equation}
c_{a,b} = \int_0^T(s'_a(t) \cdot s'_b(t))dt.
\label{eq:velcorr}
\end{equation}
We omit to compute correlation over different displacements
in time as we are only interested in it as a measure of
simultaneity. To ease later interpretation of the correlation measure,
we normalized all non-zero elements $\hat{c}_{a,b} =
\frac{c_{a,b}}{\sqrt{c_{a,a}c_{b,b}}}$. The off-diagonal elements will be valued in $[-1,1]$ depending on whether
two sliders $a$ and $b$ have been moved into opposite or same direction. A value of $1$ with this normalization corresponds to completely identical movement, and is also found on the diagonal for elements $\hat{c}_{a,a}$ of sliders that have been moved.

\begin{figure*}[h]
\centering
\subfigure[total time]{\includegraphics[width=8cm]{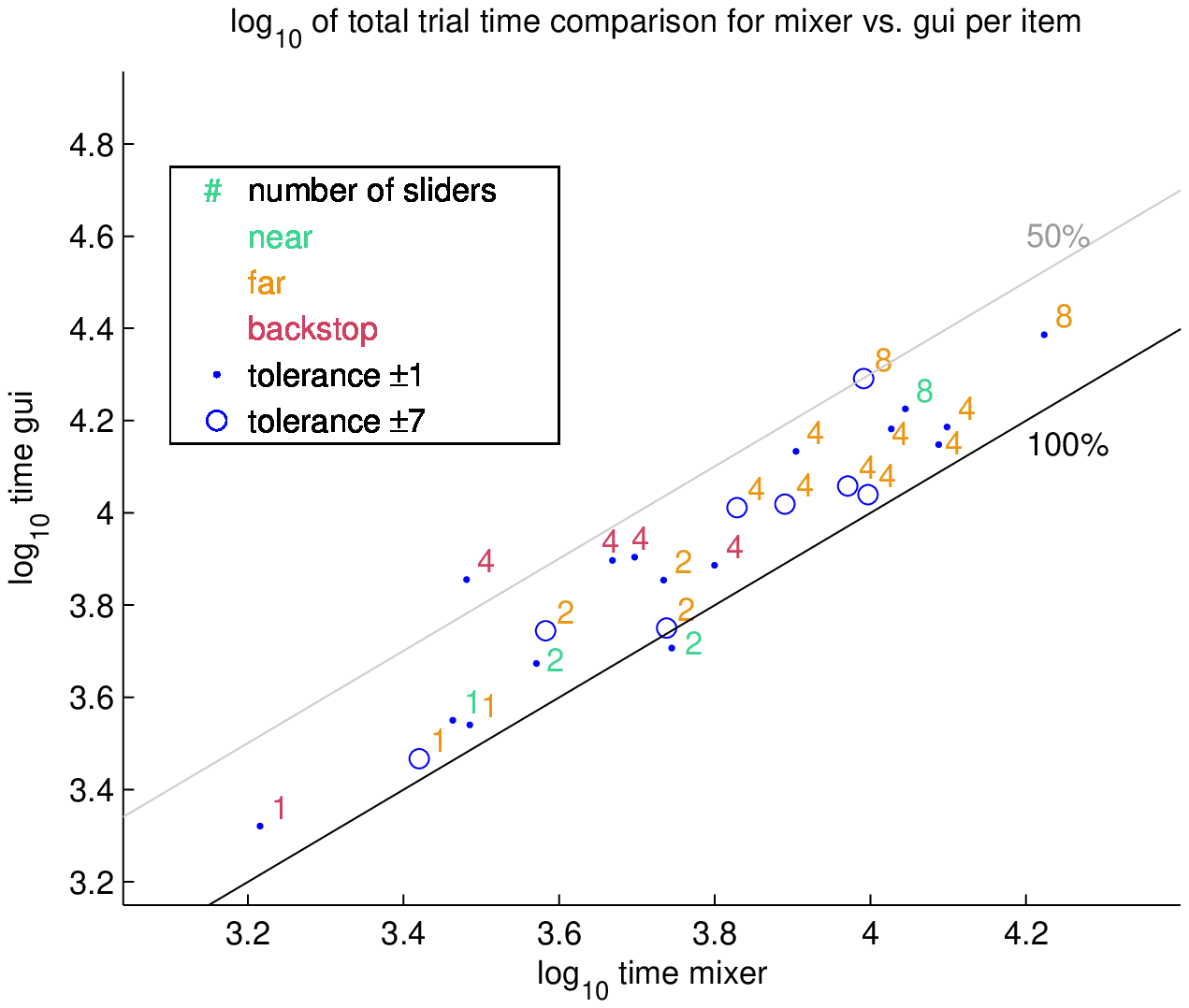}}
\subfigure[acquisition time]{\includegraphics[width=8cm]{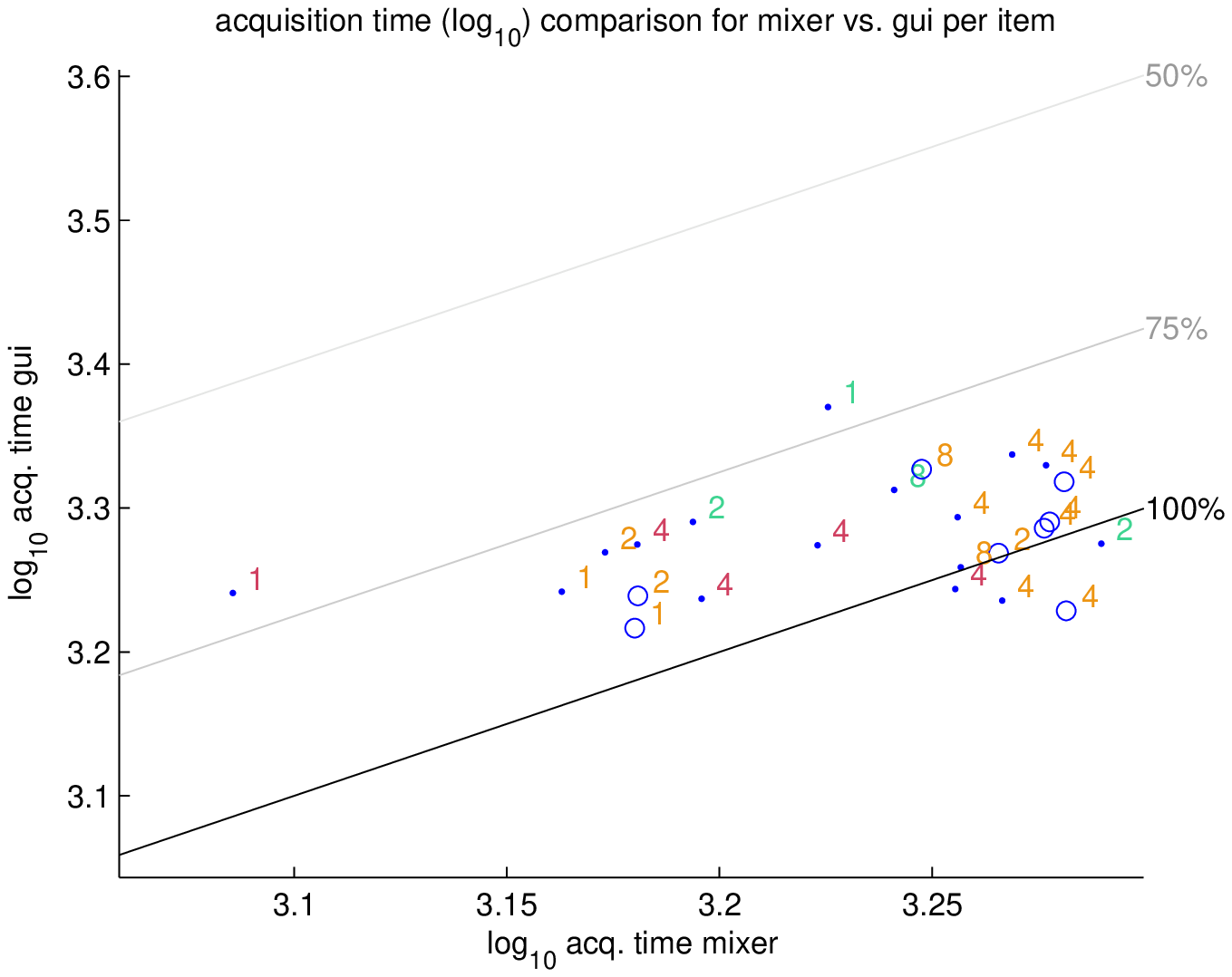}}
\subfigure[between time]{\includegraphics[width=8cm]{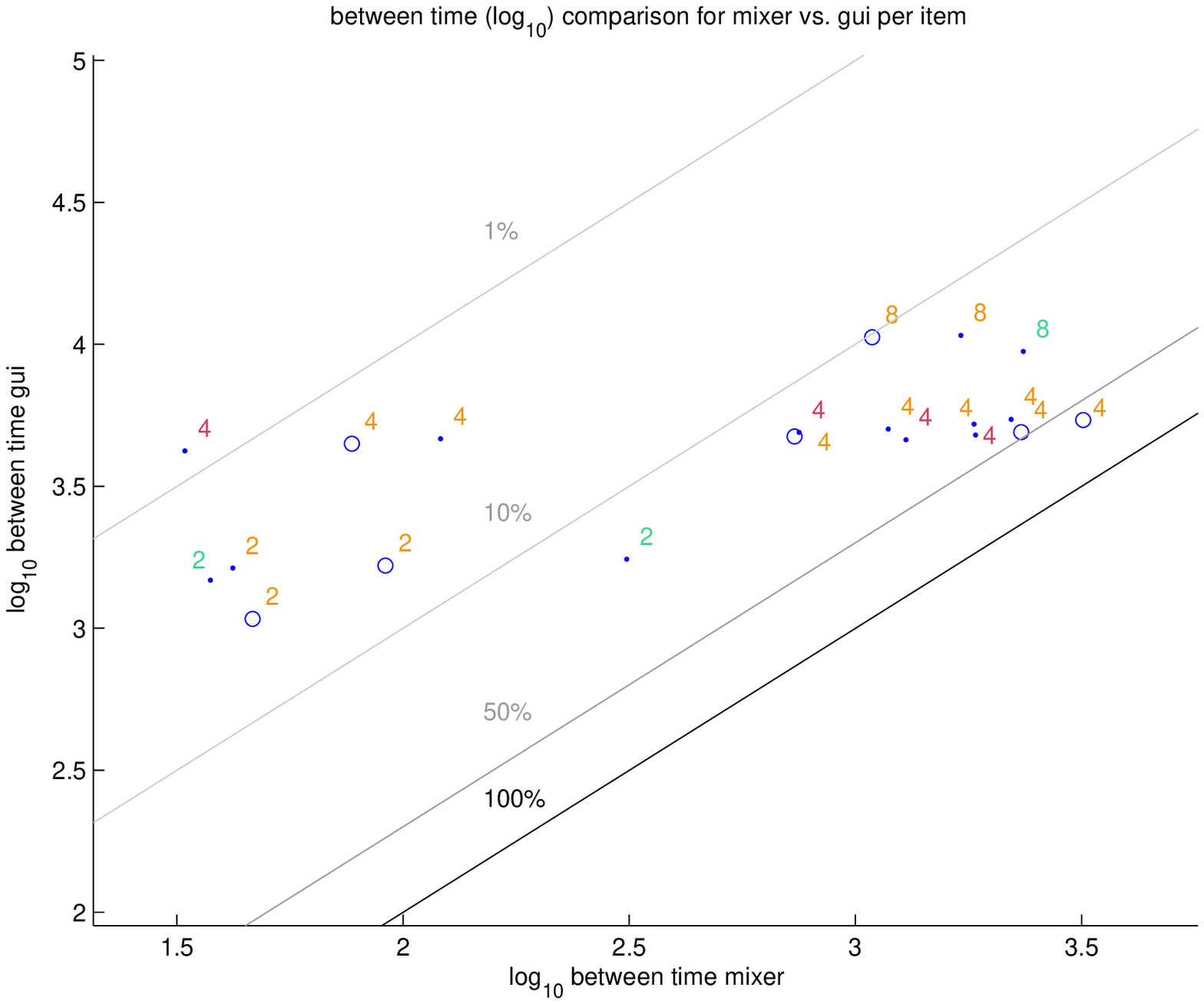}}
\subfigure[manipulation time]{\includegraphics[width=8cm]{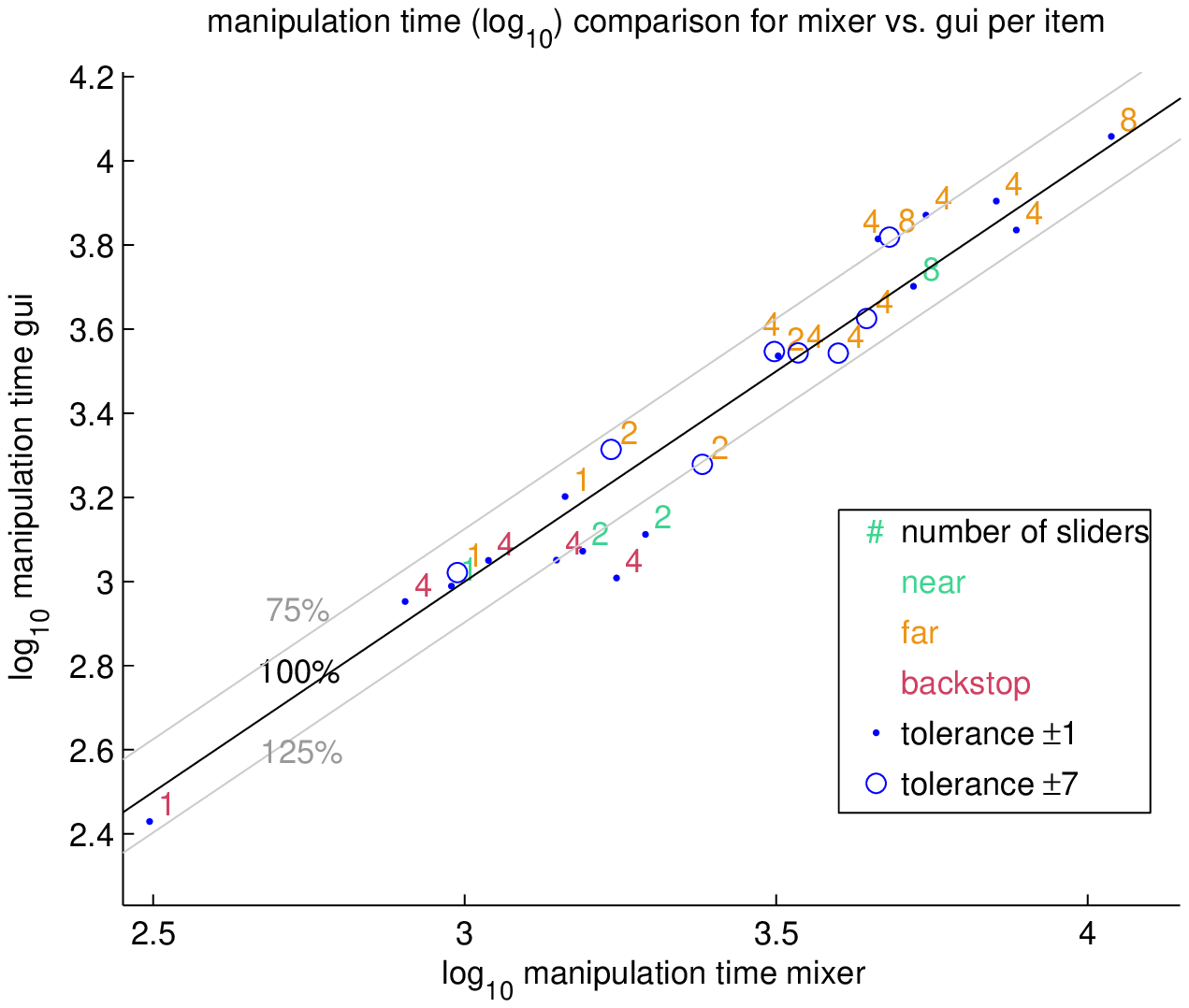}}
\caption{\small Time comparisons of trial times for mouse vs. mixer control. Each point represents the mean times taken for the items of \FG{fig:items}. The diagonal lines indicate ratios of the time taken by the mixer as compared to the mouse, e.g. $50\%$ means using the mixer took half the time of the mouse. Horizontal and vertical position are the $log_{10}$-scaled times using mouse and mixer, respectively. The label numbers indicate number of sliders, 
their colour indicates distance of the values and point  vs. circle shape indicates tight or loose precision, respectively.\label{fig:itemcomp}}
\end{figure*}
The resulting correlation matrix $C$ contains correlations among pairs
of sliders and may be used for analysing slider adjustment strategies
for a set of trials. For that we averaged over all trials for an item/input combination as in the
right column of \FG{fig:ivs}. In the mixer case (top) the off-diagonal
positive correlation (orange squares) indicates simultaneous slider
movement as positive correlation between sliders 6, 8, and 7, but not
between 6 and 7. In the mouse adjustment row at the bottom no
correlation occurs. This is not surprising, since the mouse can only control one slider at a time. 
The absolute maximum in the upper triangle of the slider-slider correlation matrix is reported as a measure of manipulation simultaneity for the item.

\section{Results} \label{sec:results} 
\subsection{Comparing Mixer vs. Mouse interaction}
The recorded timing data were distributed log-normally. Consequently, all analysis of variance was performed on the log of the times and effect sizes are reported as percentage changes in geometric mean.
Quantitative results are reported using both graphs (\FIG{fig:itemcomp}) and the multivariate form of repeated-measures analysis of variance (ANOVA).  All ANOVAs were performed using the means of the log times for each participant, for 11 denominator degrees of freedom.

\para{Timing}
The mixer was 24\% faster than the mouse for total time ($p~<~.001$), 10\% faster for acquisition time ($p~<~.042$), not significantly faster for manipulation time, and 81\% faster for between time ($p~<~.002$).  
\FG{fig:itemcomp} shows the length of different activity intervals (total time, between break times, manipulation time) for each of the 24 different items used in the study. The means used here are computed for the trials of one item over all participants for blocks 2 and 3.
This illustrates the following differences:  For total time, acquisition time, and between time, virtually all points lie above the main diagonal, indicating that the times for the mouse were larger than those for the mixer.  The between times for the mixer are virtually all 50\% or less than between times for the mouse.
Overall, $7.5\%$ of the non-manipulation time (acquisition time + between time) has been spent adjusting erroneous mixer sliders. Using mouse and GUI sliders this rate is lower at $1.1\%$.

\begin{figure}[hb]
\centering
\includegraphics[width=13cm]{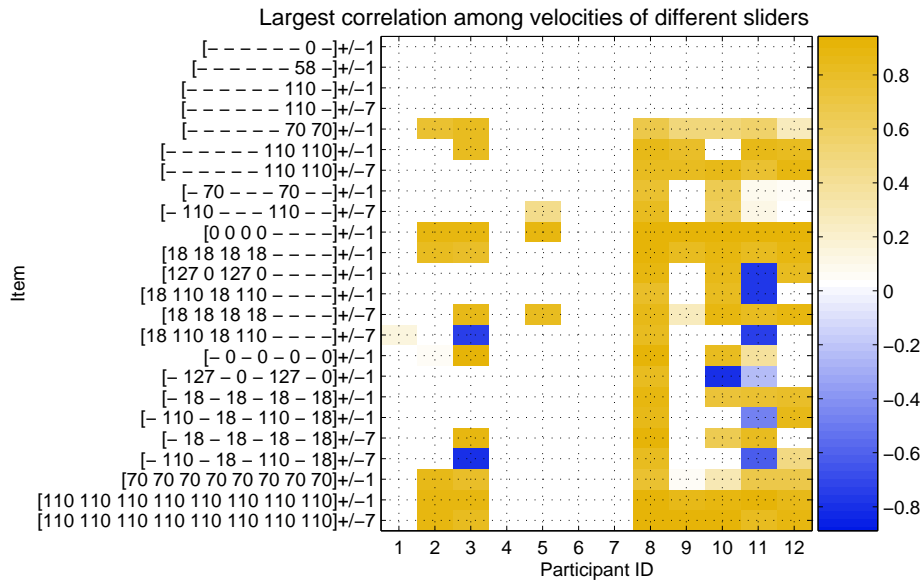}
\caption{
Simultaneous manipulation of sliders as indicated by the maximum (in absolute value) of the normalized slider-slider velocity cross-correlations of \EQ{eq:velcorr}. Simultaneity varies for different items and participants.\label{fig:simul}}
\end{figure}
\para{Simultaneity}
Our analysis of simultaneous adjustments is summarized in Fig.~\ref{fig:simul}. Here, we consider the maximum magnitude
correlation between different sliders (that is off-diagonal elements of $C$ only). Each colored pixel in the grid represents a set of mixer trials of a participant for a given item. 
One can observe the effects of different strategies. 
Three participants never made use of any
simultaneous adjustments. Three others even moved sliders
simultaneously into opposite directions, as indicated by the blue
cells. 
Items with target values that required adjustment into the
same direction on adjacent sliders caused more simultaneity than items requiring movement of 
non-adjacent sliders.

\subsection{Qualitative Data}
\FG{fig:workloadAvgs} shows the weighted results from the NASA-TLX workload questionnaires employed in our study.  Qualitatively, one can see that the mixing board was rated less demanding on all six factors.  The most significant difference reported concerns the physical demand of the two interfaces, wherein the graphical controls were rated more than twice as demanding.  Several participants reported on two reasons that may describe this conclusion: Holding the mouse over time began to make their hands sore, and they had to extend their arm more with the mouse to make an adjustment.
\begin{figure}[h]
\centering
\includegraphics[width=14cm]{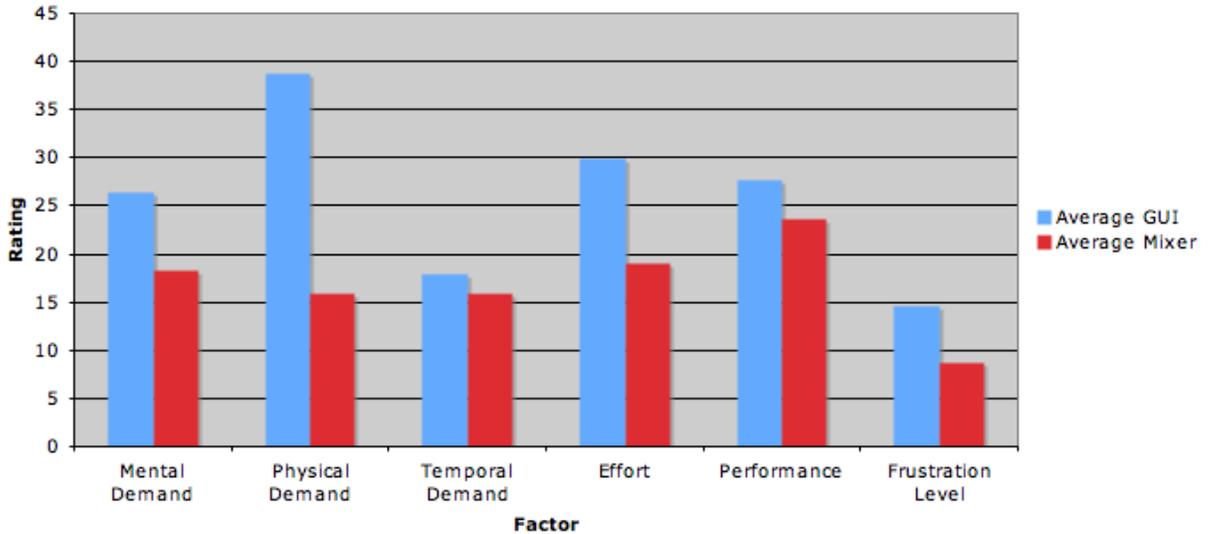}
\caption{Averaged weighted NASA-TLX workload scores\label{fig:workloadAvgs}}
\end{figure}

In the opinion questionnaires, all of the participants stated that they preferred the mixing board to the screen controls, with three participants explicitly stating that the mixer was more enjoyable or the mouse more boring.  Most of the participants also used both hands when adjusting physical sliders on some trials and/or reported bimanual input as an advantage particular to the mixing board.  Three participants reported that the mixing board was more precise than the graphical controls, and just as many reported that the physical slider knobs were easy to grab.  The main disadvantage to the mixing board reported is that users had to frequently move their eyes from the screen to the board to home their hands on to the correct knobs.  We address this issue further in our discussion.

\section{Discussion and Future Directions}

To summarize the main results of the previous section,
the overall time to complete a trial with mixer interaction is $78\%$
of mouse/GUI pointer based interaction. For items including extremal
values (backstop) the rate can go as low as $72\%$, corresponding to a
gain in speed of close to 40\% when using the mixer. The overall
distribution is shown in \FG{fig:itemcomp}a.

Determining activity intervals, as described in \SC{sec:ivdecomp}, allowed to distinguish the different
periods of non-interaction into initial acquisition time and
in-between acquisition time. The latter is significantly ($p<.001$)
dependent on the input device. Remarkably, the manipulation time
showed no significant dependence as can also be seen by inspecting
\FG{fig:itemcomp}d.

A trial task was displayed on
the screen while the manipulation devices - mixer and mouse to its right - were
placed on the table in front of the screen. Mouse interaction works
via an intermediate pointer or cursor representation on the screen,
and thus in the same perceptual space the task is presented. The
mixing board has its manipulation elements in physical space. These
need to be found using the hands and potentially moving the eye gaze
over to the board. To deal with this situation several participants
adopted the strategy of keeping their gaze on the screen while blindly
operating the sliders and occasionally touching a wrong one. Most of
these erroneous manipulations were short term mini-manipulations in
search of the right one to adjust. As reported in the analysis, the
overall amount of erroneous adjustments in the mixer manipulation made
only a small difference to the {\it between time} and were higher than
the mouse/GUI. Overall, the between time (containing the {\em error time})
was much shorter for the mixer than for the mouse/GUI based
interaction.

\para{Applications}
The shortened time of interaction and the ability to make simultaneous
adjustements make the mixing board a suitable device for user-guided
search over multi-parameter spaces as it occurs in a variety of
application settings. For example, the sliders could be used to apply ratings
of relevance, influencing an algorithms notion of 'good' search
results~\cite{Bergner:2011:paraglide,Torsney-Weir:2011:tuner,Bruckner:2010:opteffect}.

Alternatively, parameters may be influenced directly as it would be
useful to explore a design space or to configure visualizations. The user
may determine a most promising direction of exploration by learning about
cross-dependencies among different parameters. 

With separate adjustments, such as done via the mouse, the direction
of the steps in which one can move through parameter space will always
be along parameter axes. With the simultaneous adjustment of the mixer
sliders it is also possible to progress along diagonal directions in the planes spanned by any pair of parameter axes. 
This may be helpful for steering more directly towards
optimal configurations.

Another advantage for the mixer in such exploration settings is the
fact that it frees up screen real estate. With the separate
interaction device, the entire display can be used to show the data in
more detail.

In our experiments we have used the motor control of the sliders only
to ensure a clearly defined initial state for each task. However,
depending on the application setting there is more to be gained from
this feature. In particular, parameter dependencies and constraints
could be expressed to limit the user's search of feasible regions, or enable the user to push its boundaries.  
In a dynamic query range selection \cite{Crider:2007:MBI} we have
enforced the constraint that the upper bound slider may always be at a
value greater or equal to the lower bound slider. 

\para{Further directions}
One could also investigate the integration of a
separate slider adjustment as provided by the mixing board with more
advanced mouse/GUI interactions~\cite{van:1993:hyperslice,Tweedie:1998:prosection}. In \cite{Bergner:2005:FSVR} we have
devised a 2D slider embedding allowing for single click and drag
control of multi-dimensional weights. They were used to
determine a mixture of light sources, but an extension to more general application
settings is possible~\cite{Kilian:2007:shapespace,Ovsjanikov:2011:shapexplore}.

Another comparison could consider touch screen interfaces, such as TouchOSC\footnote{\url{http://hexler.net/software/touchosc}}, in relation to tactile or haptic ones, like a physical mixing board.
In this direction, a first version of this report from 2008 has already inspired follow-up work \cite{Swindells:2009:mixgaze} that showed a significantly improved visual focus on the task display, when physical sliders are used instead of GUI controls, which did not occur when using touch screen sliders placed in the same prior location of the physical sliders on a table top display. While their study did not obtain the timing improvements documented in this report, the improved eye gaze focus provides further indication and a possible explanation of the reduced cognitive load when using the mixer. 

The items used in our current study are all fairly structured and contain several identical values. 
Further studies could consider more different value combinations, e.g. giving target points that lie in certain specifically shaped neighbourhoods around the origin. Different directions of the required steering could be considered, as well as measures of how exhaustively a user explored the feasible options. A readily available binding in our \sname{paraglide} system could provide the technical setup for such a study~\cite{Bergner:2011:paraglide}.

In summary, the shown characteristics portrayed mixing boards as a good fit for the purpose of a detailed multi-parameter control. Easy and relatively cheap integration of such a device into an application is possible. Its strengths should particularly hold in settings where screen space is precious, and undivided visual focus is crucial.

\section{Conclusion}
The comparative study described in this paper revealed a significant
advantage of the mixing board input over a mouse/GUI based input
reducing the overall time required to complete the multi-parameter
adjustment tasks, while also reducing the cognitive load perceived by the participants.

\para{Methodological contributions}
Beyond inspecting recorded total time we have
conducted an analysis of motion path data. This enabled the decomposition of
the overall interaction time into different activity intervals. Aside
from providing a more accurate method for data cleaning, this also
revealed a crucial effect present in the recorded data. The pure
manipuation time was independent of the input device. The significant
difference favoring the mixer in overall timing resulted solely from
improved timing when reaching for the controls, namely acquisition
time and between time.  
To investigate simultaneity of adjustments we devised a velocity
correlation measure. 

\para{Implications}
Our study highlights the potential for a mixing board to augment 
mouse control in applications where parameters need to be adjusted.  While
such applications may not have a graphical slider control,
or an analogous widget (such as a knob), the metaphor of
adjusting discrete or continuous values can typically be mapped to a
slider.  In such cases, it is advantageous to assign different
parameters to their own sliders, allowing for a spatially-multiplexed
interaction.  
The constrained motions of the mixing board sliders allowed us to identify basic interaction patterns for different types of items. Our findings indicate that in order to stimulate simultaneous slider movements, parameters that 
require correlated adjustments are best placed next to each other.
This could provide a convenient method of exploration in the direction of parameter axes and along the diagonals in low-dimensional subspaces (e.g. the planes spanned by any two axes).


\appendix
\section*{Appendix}

\section{Setup} \label{sec:setup}

All experimental software ran on a hyper-threaded dual-core Apple PowerPC ($4\times2.5$ GHz) with 8 GB RAM running MacOS 10.5.  The graphical interface was displayed in a rectangular area on a 20 inch TFT monitor.  The machine was connected to the Internet, but did not have any other applications running---there was no contention for the processor.  

\begin{wrapfigure}{r}{0.4\textwidth}
\begin{center}
\includegraphics[width=6cm]{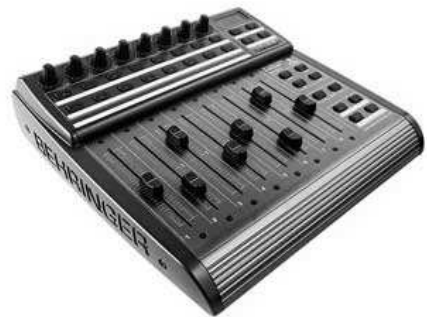}
\caption{The BCF2000 Mixing Board. 
\label{fig:bcf2000}}
\end{center}
\end{wrapfigure}
When deciding on which particular mixing board to acquire, our choices came down to two options: The BCF2000 (see \FIG{fig:bcf2000}), and the Evolution UC33e, both are easily connected to a computer through a USB or MIDI port.  
A key feature of digital MIDI mixing boards is that they are platform independent and communicate with a standard message protocol, which all major operating systems can understand.  Our choice of Operating System is arbitrary, and does not effect the software we developed, which has also been run on Windows and Linux Operating Systems.

We opted for the BCF2000 as, in addition to all the features shared between the two mixers it has {\em motorized sliders}. This allows to send messages to the mixing board and adjust the physical position of the sliders to synchronize them with their graphical counterparts, and enforce constraints on the sliders by only allowing them to be in certain positions.  Further, it allows the board to be used as a type of output device that conveys data values by dynamically updating the positions of the sliders.

\para{Software}
To communicate with the mixing board, we used RtMidi~\cite{Scavone:2005:rtmidi}, an open source and multiplatform C++ API for communicating with MIDI devices. 
On top of this API, using the Qt GUI toolkit, we developed an application that provides a graphical representation of the device's controls (see \FIG{fig:midTrial}).

To generate trial values, we used an Experimental Driver program that facilitates the formulation of trial data in an XML file. This allows for a simple means of controlling the order of the trials in the experiment.  Further, it enabled us to randomize the trial types that our participants would receive within a set block of experiments. This driver included a mechanism to receive results from our client application and to organize them in logs for further analysis.

\bibliographystyle{acmtrans}
\bibliography{main}

\begin{thebibliography}{}

\bibitem[\protect\citeauthoryear{Bergner, M{\"o}ller, Tory, and Drew}{Bergner
  et~al\mbox{.}}{2005}]{Bergner:2005:FSVR}
{\sc Bergner, S.}, {\sc M{\"o}ller, T.}, {\sc Tory, M.}, {\sc and} {\sc Drew,
  M.~S.} 2005.
\newblock A practical approach to spectral volume rendering.
\newblock {\em IEEE Trans. on Vis. and Comp. Graphics\/}~{\em 11,\/}~2,
  207--216.

\bibitem[\protect\citeauthoryear{Bergner, Sedlmair, Nabi-Abdolyousefi, Saad,
  and M{\"o}ller}{Bergner et~al\mbox{.}}{2011}]{Bergner:2011:paraglide}
{\sc Bergner, S.}, {\sc Sedlmair, M.}, {\sc Nabi-Abdolyousefi, S.}, {\sc Saad,
  A.}, {\sc and} {\sc M{\"o}ller, T.} 2011.
\newblock {Paraglide: Interactive Parameter Space Partitioning for Computer
  Simulations}.
\newblock Submitted to IEEE Trans. on Vis. and Comp. Graphics, Sep. 2011.

\bibitem[\protect\citeauthoryear{Bruckner and M{\"o}ller}{Bruckner and
  M{\"o}ller}{2010}]{Bruckner:2010:opteffect}
{\sc Bruckner, S.} {\sc and} {\sc M{\"o}ller, T.} 2010.
\newblock Result-driven exploration of simulation parameter spaces for visual
  effects design.
\newblock {\em IEEE Trans. on Vis. and Comp. Graph. (Proc. Vis. / Info. Vis.
  2010)\/}~{\em 16}, 1467--1475.

\bibitem[\protect\citeauthoryear{Buxton and Myers}{Buxton and
  Myers}{1986}]{buxton86}
{\sc Buxton, W.} {\sc and} {\sc Myers, B.} 1986.
\newblock A study in two-handed input.
\newblock In {\em Proc. CHI 1986}. ACM Press, Boston, MA, 321--326.

\bibitem[\protect\citeauthoryear{Chipman, Bederson, and Golbeck}{Chipman
  et~al\mbox{.}}{2004}]{chipman04}
{\sc Chipman, L.~E.}, {\sc Bederson, B.~B.}, {\sc and} {\sc Golbeck, J.~A.}
  2004.
\newblock {SlideBar: analysis of a linear input device}.
\newblock {\em Behaviour and Information Technology\/}~{\em 23,\/}~1, 1--9.

\bibitem[\protect\citeauthoryear{Crider, Bergner, Smyth, Kirkpatrick, and
  M{\"o}ller}{Crider et~al\mbox{.}}{2007}]{Crider:2007:MBI}
{\sc Crider, M.}, {\sc Bergner, S.}, {\sc Smyth, T.}, {\sc Kirkpatrick, A.},
  {\sc and} {\sc M{\"o}ller, T.} 2007.
\newblock A mixing board interface for graphics and visualization applications.
\newblock In {\em Proc. Graphics Interface 2007}. ACM, Montreal, QC, 87--94.

\bibitem[\protect\citeauthoryear{Faisal, Cairns, and Craft}{Faisal
  et~al\mbox{.}}{2005}]{Faisal05}
{\sc Faisal, S.}, {\sc Cairns, P.}, {\sc and} {\sc Craft, B.} 2005.
\newblock Infoviz experience enhancement through mediated interaction.
\newblock In {\em ICMI'05 Workshop on Multimodal Interaction for the
  Visualisation and Exploration of Scientific Data}. ITC/ACM, Trento, IT, 3--9.

\bibitem[\protect\citeauthoryear{Fitzmaurice and Buxton}{Fitzmaurice and
  Buxton}{1997}]{Fitzmaurice97}
{\sc Fitzmaurice, G.~W.} {\sc and} {\sc Buxton, W.} 1997.
\newblock An empirical evaluation of graspable user interfaces: towards
  specialized, space-multiplexed input.
\newblock In {\em Proc. CHI 1997}. ACM Press, New York, NY, 43--50.

\bibitem[\protect\citeauthoryear{Hart and Staveland}{Hart and
  Staveland}{1988}]{Hart:1988:TLX}
{\sc Hart, S.} {\sc and} {\sc Staveland, L.} 1988.
\newblock {Development of NASA-TLX (Task Load Index): Results of empirical and
  theoretical research}.
\newblock {\em Human mental workload\/}~{\em 1}, 139--183.

\bibitem[\protect\citeauthoryear{Hinckley, Pausch, Goble, and Kassell}{Hinckley
  et~al\mbox{.}}{1994}]{Hinckley94}
{\sc Hinckley, K.}, {\sc Pausch, R.}, {\sc Goble, J.~C.}, {\sc and} {\sc
  Kassell, N.~F.} 1994.
\newblock Passive real-world interface props for neurosurgical visualization.
\newblock In {\em Proc. CHI 1994}. ACM Press, Boston, MA, 452--458.

\bibitem[\protect\citeauthoryear{Hunt and Kirk}{Hunt and Kirk}{1999}]{HuntK99}
{\sc Hunt, A.} {\sc and} {\sc Kirk, R.} 1999.
\newblock Radical user interfaces for real-time control.
\newblock In {\em EUROMICRO}. IEEE Computer Society, Milan, IT, 2006--2012.

\bibitem[\protect\citeauthoryear{Jacob, Sibert, McFarlane, and
  M.~Preston~Mullen}{Jacob et~al\mbox{.}}{1994}]{Jacob:1994:integral}
{\sc Jacob, R. J.~K.}, {\sc Sibert, L.~E.}, {\sc McFarlane, D.~C.}, {\sc and}
  {\sc M.~Preston~Mullen, J.} 1994.
\newblock Integrality and separability of input devices.
\newblock {\em ACM Transactions on Computer-Human Interaction\/}~{\em 1,\/}~1,
  3--26.

\bibitem[\protect\citeauthoryear{Kabbash, Buxton, and Sellen}{Kabbash
  et~al\mbox{.}}{1994}]{Kabbash94}
{\sc Kabbash, P.}, {\sc Buxton, W.}, {\sc and} {\sc Sellen, A.} 1994.
\newblock Two-handed input in a compound task.
\newblock In {\em Proc. CHI 1994}. ACM Press, Boston, Massachusetts, 417--423.

\bibitem[\protect\citeauthoryear{Kilian, Mitra, and Pottmann}{Kilian
  et~al\mbox{.}}{2007}]{Kilian:2007:shapespace}
{\sc Kilian, M.}, {\sc Mitra, N.~J.}, {\sc and} {\sc Pottmann, H.} 2007.
\newblock Geometric modeling in shape space.
\newblock {\em ACM Transactions on Graphics\/}~{\em 26,\/}~3, 1--8.

\bibitem[\protect\citeauthoryear{Konieczny, Shimizu, Meyer, and
  Colucci}{Konieczny et~al\mbox{.}}{2005}]{KoniecznySMC05}
{\sc Konieczny, J.}, {\sc Shimizu, C.}, {\sc Meyer, G.~W.}, {\sc and} {\sc
  Colucci, D.} 2005.
\newblock A handheld flexible display system.
\newblock In {\em Proc. VIS 2005}. IEEE Computer Society Press, Minneapolis,
  MN, 75.

\bibitem[\protect\citeauthoryear{McKelvin, Nestande, Valdez, Yee, Back, and
  Harrison}{McKelvin et~al\mbox{.}}{2003}]{McKelvin03}
{\sc McKelvin, M.}, {\sc Nestande, R.}, {\sc Valdez, L.}, {\sc Yee, K.-P.},
  {\sc Back, M.}, {\sc and} {\sc Harrison, S.} 2003.
\newblock Seismospin: a physical instrument for digital data.
\newblock In {\em CHI 2003: extended abstracts on Human factors in computing
  systems}. ACM Press, Ft. Lauderdale, Florida, 832--833.

\bibitem[\protect\citeauthoryear{Ovsjanikov, Li, Guibas, and Mitra}{Ovsjanikov
  et~al\mbox{.}}{2011}]{Ovsjanikov:2011:shapexplore}
{\sc Ovsjanikov, M.}, {\sc Li, W.}, {\sc Guibas, L.}, {\sc and} {\sc Mitra,
  N.~J.} 2011.
\newblock Exploration of continuous variability in collections of 3d shapes.
\newblock {\em ACM Trans. on Graphics\/}~{\em 30,\/}~4, to appear.

\bibitem[\protect\citeauthoryear{Rheingans}{Rheingans}{1992}]{Rheingans92}
{\sc Rheingans, P.} 1992.
\newblock Color, change, and control for quantitative data display.
\newblock In {\em Proc. VIS 1992}. IEEE Comp. Soc. Press, Boston, MA, 252--259.

\bibitem[\protect\citeauthoryear{Scavone and Cook}{Scavone and
  Cook}{2005}]{Scavone:2005:rtmidi}
{\sc Scavone, G.} {\sc and} {\sc Cook, P.} 2005.
\newblock Rtmidi, rtaudio, and a synthesis toolkit (stk) update.
\newblock In {\em In Proc. of the Intl. Comp. Music Conf.} Citeseer, Spain.

\bibitem[\protect\citeauthoryear{Shahrokni, Jenaro, Gustafsson, Vinnberg,
  Sandsjo, and Fjeld}{Shahrokni et~al\mbox{.}}{2006}]{Shahrokni2006}
{\sc Shahrokni, A.}, {\sc Jenaro, J.}, {\sc Gustafsson, T.}, {\sc Vinnberg,
  A.}, {\sc Sandsjo, J.}, {\sc and} {\sc Fjeld, M.} 2006.
\newblock {One-dimensional force feedback slider: Going from an analogue to a
  digital platform}.
\newblock In {\em Proc. NordiCHI 2006}. ACM Press, Oslo, Norway, 453--456.

\bibitem[\protect\citeauthoryear{Swindells, Tory, and Dreezer}{Swindells
  et~al\mbox{.}}{2009}]{Swindells:2009:mixgaze}
{\sc Swindells, C.}, {\sc Tory, M.}, {\sc and} {\sc Dreezer, R.} 2009.
\newblock Comparing parameter manipulation with mouse, pen, and slider user
  interfaces.
\newblock {\em Computer Graphics Forum (Proc. EuroVis 2009)\/}~{\em 28,\/}~3
  (June), 919--926.

\bibitem[\protect\citeauthoryear{Torsney-Weir, Saad, M{\"o}ller, Weber, Hege,
  Verbavatz, and Bergner}{Torsney-Weir
  et~al\mbox{.}}{2011}]{Torsney-Weir:2011:tuner}
{\sc Torsney-Weir, T.}, {\sc Saad, A.}, {\sc M{\"o}ller, T.}, {\sc Weber, B.},
  {\sc Hege, H.-C.}, {\sc Verbavatz, J.-M.}, {\sc and} {\sc Bergner, S.} 2011.
\newblock {Tuner: Principled Parameter Finding for Image Segmentation
  Algorithms Using Visual Response Surface Exploration}.
\newblock {\em IEEE Trans. on Vis. and Comp. Graphics\/}~{\em 18,\/}~6,
  ???--???

\bibitem[\protect\citeauthoryear{Tweedie and Spence}{Tweedie and
  Spence}{1998}]{Tweedie:1998:prosection}
{\sc Tweedie, L.} {\sc and} {\sc Spence, R.} 1998.
\newblock The prosection matrix: A tool to support the interactive exploration
  of statistical models and data.
\newblock {\em Computational Statistics\/}~{\em 13,\/}~1, 65--76.

\bibitem[\protect\citeauthoryear{van Wijk and van Liere}{van Wijk and van
  Liere}{1993}]{van:1993:hyperslice}
{\sc van Wijk, J.} {\sc and} {\sc van Liere, R.} 1993.
\newblock {HyperSlice: Visualization of scalar functions of many variables}.
\newblock In {\em Proc. of 4th Conf. on Visualization'93}. IEEE Computer
  Society, San Jose, 119--125.

\end{thebibliography}

\end{document}